# Using Tidal Tails to Probe Dark Matter Halos

John Dubinski, J. Christopher Mihos,[1,2] and Lars Hernquist,[3]
Board of Studies in Astronomy and Astrophysics
University of California, Santa Cruz, CA 95064
dubinski@lick.ucsc.edu, hos@lick.ucsc.edu, lars@lick.ucsc.edu

## ABSTRACT

We use simulations of merging galaxies to explore the sensitivity of the morphology of tidal tails to variations of the halo mass distributions in the parent galaxies. Our goal is to constrain the mass of dark halos in well-known merging pairs. We concentrate on prograde encounters between equal mass galaxies which represent the best cases for creating tidal tails, but also look at systems with different relative orientations, orbital energies and mass ratios. As the mass and extent of the dark halo increase in the model galaxies, the resulting tidal tails become shorter and less massive, even under the most favorable conditions for producing these features. Our simulations imply that the observed merging galaxies with long tidal tails ($\sim 50 - 100$ kpc) such as NGC 4038/39 (the Antennae) and NGC 7252 probably have halo:disk+bulge mass ratios less than 10:1. These results conflict with the favored values of the dark halo mass of the Milky Way derived from satellite kinematics and the timing argument which give a halo:disk+bulge mass ratio of $\sim 30 : 1$. However, the lower bound of the estimated dark halo mass in the Milky Way (mass ratio $\sim 10 : 1$) is still consistent with the inferred tidal tail galaxy masses. Our results also conflict with the expectations of $\Omega = 1$ cosmologies such as CDM which predict much more massive and extended dark halos.

*Subject headings:* galaxies:interactions – galaxies:structure – dark matter – cosmology:dark matter

---

[1]Hubble Fellow

[2]Current Address: Department of Physics and Astronomy, The Johns Hopkins University, Baltimore, MD 21218

[3]Alfred P. Sloan Foundation Fellow, Presidential Faculty Fellow

## 1. Introduction

There is now considerable observational evidence for the presence of large amounts of dark matter in the Universe. On the scales of individual galaxies, the mass-to-light ratio of spirals and ellipticals is in the range $\sim$ 10-50 $(M/L)_\odot$, well in excess of the mass-to-light ratio for normal stellar populations $\sim 3$ $(M/L)_\odot$ (e.g., Binney & Tremaine 1987). On larger scales associated with groups and clusters of galaxies, the mass-to-light ratios are even higher, typically $\sim 100 - 500$ $(M/L)_\odot$. Clearly, ordinary "luminous" matter such as stars and stellar remnants cannot account for such extreme values. On the largest scales, an even higher total mass-to-light ratio for the Universe of $\sim 1500h$ ($h = H_0/100$ km/s/Mpc) is required for closure density in a critical, $\Omega = 1$ universe. Determining the amount of dark matter present in the Universe is thus a key ingredient in understanding cosmological evolution.

In principle, cosmological models can be constrained by the structure of individual galaxies, if we can infer the masses and extents of their dark matter halos. Gunn & Gott (1972) recognized that spherical infall of dissipationless matter onto a density perturbation in an Einstein-de Sitter universe would relax into an object with a density profile $\rho \propto r^{-9/4}$, similar to the inferred mass distribution around galaxies (see also Bertschinger 1985). Furthermore, if $\Omega = 1$, mass continues to accrete continuously so that halo masses grow with time as $M \sim t^{2/3}$, implying that dark halos should extend to great distances. In the case of the Milky Way, a $\rho \sim r^{-2}$ profile extrapolated from the local rotation curve reaches the background density at a radius $\sim 1$ Mpc. In contrast, in subcritical universes collapsing objects stop accreting at a redshift, $z \sim 1/\Omega$, and therefore have a steeper density profile than they would if $\Omega = 1$ (e.g., Hoffman & Shaham 1985; Zurek, Quinn, & Salmon 1988). Measurements of the properties of dark halos in cosmological N-body simulations support the predictions of the spherical accretion calculations (e.g. Zurek, Quinn, & Salmon 1988). For example, the cold dark matter (CDM) model predicts that dark halos associated with galaxies like the Milky Way should extend well beyond 200 kpc at the present time, with approximately isothermal density profiles and masses well in excess of $10^{12}$ $M_\odot$ (e.g., Navarro, Frenk, & White 1995).

The strongest observational constraints on the amount of dark matter around galaxies comes from the kinematics of spirals. The rotation curves of spiral galaxies, as inferred by optical spectra and HI linewidths, are roughly flat out to $\sim 10$ disk scale lengths, or $\sim 30 - 40$ kpc, (e.g. Rubin et al. 1980, 1982, 1985; Kent 1987), contrary to the assumption that light traces mass. The usual interpretation of this finding is that galaxies are surrounded by unseen dark halos. At large radii, the dark halo dominates the mass distribution and so a flat rotation curve implies a density profile $\rho(r) \propto r^{-2}$ and, hence, $M(r) \propto r$. The decomposition of rotation curves suggests that the mass ratio of halo to disk plus bulge matter is approximately 10:1 out to the edges of HI gas disks (e.g., Kent 1987). A further outwards extrapolation of the profile $M \propto r$ implies that halos could be much more massive, but their extents and total masses cannot be determined from rotation curves alone.

The hot, X-ray coronae around some elliptical galaxies also seems to require massive dark halos to support them (e.g., Forman, Jones, & Tucker 1985). If the gas is in hydrostatic equilibrium, the gravitational potential can be determined by analyzing the X-ray emission profile. Typically, a mass model with $\rho \propto r^{-2}$ fits the data well, consistent with the implications of spiral rotation curves, but again in most cases the total masses and extents of the halos cannot be determined unambiguously.

In an effort to probe the nature of halos on scales larger than those covered by rotation curves, Zaritsky & White (1994) have examined the distribution and kinematics of satellites around external galaxies. Since the objects in their sample were chosen to have similar luminosities and Hubble types, Zaritsky & White were able to perform an ensemble average over all of their galaxies to estimate halo masses. Within the uncertainties of projection effects, they conclude that halo masses are $\sim 1 - 2 \times 10^{12}$ M$_\odot$ for galaxies with rotation velocities similar to the Milky Way, $V_c = 220$ km s$^{-1}$.

Brainerd, Blandford & Smail (1995) have introduced a new technique to study halos which is based on weak gravitational lensing of faint background galaxies by brighter foreground galaxies.[4] Deep CCD exposures show that fainter galaxies have a tendency to be tangentially aligned around the brighter (presumably closer) galaxies in projection. By modeling the galaxy redshift distribution, Brainerd et al. show that this result can be explained by weak gravitational lensing if galaxy halos extend to at least $\sim 100$ kpc and have characteristic masses at least $\sim 10^{12}$ M$_\odot$. Their detection of the effect is marginal at present, but in principle, this method can be used to determine the masses and extents of halos without any dynamical modeling.

Our own Galaxy provides a unique opportunity to study dark halos. The rotation curve of the Milky Way is nearly flat out to a distance of at least 20 kpc (e.g. Fich & Tremaine 1991) making it similar to external galaxies. Beyond that, the mass distribution has been estimated using various tracers of the gravitational potential which are inaccessible in external galaxies. High velocity stars in the solar neighborhood set a lower limit on the local escape velocity. The analysis by Leonard & Tremaine (1990) gives a lower limit to the mass of the Milky Way of $4.6 \times 10^{11}$ M$_\odot$. Further out, the kinematics of distant globular clusters and dwarf galaxies trace the mass profile in the outer galaxy from 50 to 200 kpc (Hartwick & Sargent 1978; Lynden-Bell, Cannon & Godwin 1983; Little & Tremaine 1987). The most recent estimates based on this approach (e.g., Zaritsky et al. 1989; Kochanek 1995) suggest that the Milky Way's mass is $\sim 1 \times 10^{12}$ M$_\odot$ within a radius of 100 kpc. Unfortunately, this method is particularly sensitive to whether or not the dwarf Leo I is included in the sample. Leo I lies at a large distance from the Galactic center ($r = 220$ kpc) and has a high radial velocity ($v = 177$ km s$^{-1}$); removing it from the analysis lowers the estimated mass within 100 kpc by $\sim 50\%$ in Kochanek's (1995) analysis.

---

[4]The work is similar to current efforts to map the dark matter distribution around clusters using weak lensing (e.g., Kaiser et al. 1995).

By modeling the orbital dynamics of the Large and Small Magellanic Clouds, and by requiring that the Magellanic Stream originated from tidal stripping, Lin et al. (1995) estimate the mass of the Milky Way. At the distance of the LMC (55 kpc), their estimated mass of $5 \times 10^{11}$ M$_\odot$ is consistent with the mass interior to 50 kpc inferred from satellite kinematics.

Finally, the mass of the Local Group can be estimated from the timing argument (Kahn & Woltjer 1959), which is founded on the assumption that M31 and the Milky Way are bound and are currently on the return portion of a nearly radial orbit which originated at the Big Bang. From the current separation and relative radial velocity of the two galaxies, the derived total mass is between $3.5 \times 10^{12}$ and $5.6 \times 10^{12}$ M$_\odot$ for Universe ages between 10 and 20 Gyr (with the higher mass being associated with the lower age). Andromeda has a larger circular velocity and disk scale length than the Milky Way, suggesting that it is $\sim 2$ times the mass of the Galaxy (e.g., Raychaudury & Lynden-Bell 1989). This sets the mass of the Milky Way at $\approx 1 - 2 \times 10^{12}$ M$_\odot$ depending on the age of the Universe, consistent with the mass derived from satellite kinematics (Zaritsky et al. 1989; Kochanek 1995). These total masses imply a halo:disk+bulge mass ratio of 15-30:1 using the estimated disk+bulge mass of $\sim 6 \times 10^{10}$ M$_\odot$ (e.g., Bahcall & Soneira 1980).

In this paper, we seek to probe the structure of dark halos in the outer regions of galaxies in an independent manner, by examining the structure of the often long tidal tails associated with merging galaxies. Among the well-known examples of ongoing mergers, several have prominent tidal tails (Arp 1966) – e.g. NGC 4038/39 ("The Antennae"), NGC 7252, Arp 193, and Arp 243, to name a few. In projection, the tails in these objects extend out to $\sim 10 - 20$ disk scale lengths from the merging pair (Schweizer 1982; Schombert, Wallin & Struck-Marcell 1990; Hibbard 1994). The tails probably extend to even greater physical distances in three dimensions, perhaps $\sim 20 - 40$ scale lengths. The Superantennae (IRAS 19254-7245) is an extreme case in which the tails span $\sim 350$ kpc from tip to tip (Mirabel, Lutz, & Maza 1991; Colina, Lipari, & Macchetto 1991).

The origin of the thin tails extending from interacting galaxies is well understood.

Toomre & Toomre (1972) and Wright (1972) demonstrated that thin tails could be produced by tidal forces during a close encounter of two disk galaxies, finally laying to rest lingering doubts about the gravitational origin of these features. However, these calculations modeled the potential of the galaxies as point masses, and did not explore variations in tidal tail morphology due to the presence of dark halos. Faber and Gallagher (1979) later speculated that the lengths of these tidal features might be used to constrain the mass distributions of the outer regions of galaxies. In principle, tidal tails trace the orbit of particles ejected from the disk at some high velocity acquired in the interaction and so trace the potential of the galaxies at large radii. In practice, the strength of the perturbation depends on many factors including the perigalactic separation and velocity, disk orientation, and galaxy mass ratio. An extensive survey of possible galaxy collisions is therefore required to determine this effects.

Simulations employing self-consistent galaxies have put such arguments on a more quantitative footing and moved towards attaining this goal. White (1982) and Negroponte & White (1983) showed that tidal tails are readily produced in self-consistent mergers. Because of computational limitations, however, they were unable to explore parameter space fully, and noted that galaxies with very massive halos might have difficulty producing tails, for the following reasons. When galaxies with massive halos collide, their encounter velocities will be high since the pair falls into a a deep potential well during the encounter. A higher velocity interaction can detune the resonance between the orbital angular frequency and the internal angular frequency of the disk stars, which is needed to produce tidal tails. Moreover, the material forming the tails would have a deeper (and perhaps steeper) potential well to climb while being ejected from each galaxy. In principle, both effects can reduce the masses and lengths of the tails.

Barnes (1988) investigated this issue by simulating collisions of pairs of identical galaxies, each having halo:disk+bulge mass ratios in the sequence 0:1, 4:1, 8:1. Although there seemed to be evidence that encounters of galaxies with more extended halos produce less massive and shorter tails, Barnes concluded that tails are relatively easy to make. However, according to observational and theoretical prejudice, appropriate halo:disk+bulge mass ratios for galaxies may well be considerably larger than those employed by Barnes.

In this paper, we revisit the issue of the sensitivity of tidal tail formation in mergers to the masses and extents of dark halos, but consider larger halos than used in earlier studies. In §2, we describe galaxy models having halo:disk+bulge mass ratios of 4:1, 8:1, 16:1, and 30:1, but similar rotation curves within 5 disk scale lengths, with the more massive halos extending to larger radii. In §3, we describe zero energy, prograde encounters of galaxies in which the impact parameter is varied. Disks suffer the strongest tidal response during prograde collisions, so this geometry should be optimal for producing tidal tails. Our simulations demonstrate that long tidal tails are difficult to produce in mergers between galaxies with the most massive halos even under these optimal conditions. We elaborate this point further in §4, by investigating variations in orbital energy, disk inclinations, and the mass ratio of the progenitors, using NGC 4038/39 and the Local Group as test cases. Finally, we consider the implications of our results for the structure of galaxies and cosmology in general in §5.

## 2. Numerical Methods

### 2.1. Galaxy Models

Our goal is to use tidal tails to constrain the amount of dark matter surrounding galaxies at radii larger than those which can be probed by optical or HI rotation curves. Accordingly, we wish to compare mergers of galaxies whose rotation curves are similar in their inner regions, where the mass distribution is relatively well-constrained, but which differ in the total masses and extents of their dark matter halos. Previous techniques for constructing model galaxies (e.g., Barnes 1988; Hernquist 1993a) are not well-suited for this task.

Instead, we employ the methods for constructing equilibrium model galaxies developed recently by Kuijken & Dubinski (1995). In particular, they present a set of galaxy models with different halo masses and extents, all of which reproduce the observed rotation curve of the Milky Way out to 5 disk scale lengths ($5R_d$). In dimensionless units, the models have $R_d = 1.0$, rotation velocity $v_c(R_d) \approx 1.0$, a disk mass, $M_d = 0.82$, and a bulge mass $M_b = 0.42$. Scaling to values appropriate for the Milky Way, these values correspond to $R_d = 4.0$ kpc, $V_c = 220$ km s$^{-1}$, $M_d = 4.4 \times 10^{10} M_\odot$, and $M_b = 2.3 \times 10^{10} M_\odot$. The halos vary in mass, yielding halo:disk+bulge mass ratios between 4:1 and 30:1, and vary in radial extent from $\approx 22$ to 73 disk scale lengths (see Table 1). The rotation curves of these models, shown in Figure 1, are nearly identical within $5R_d$, differing significantly only at large radii.

The bulge and halo components of these galaxies are derived from lowered Evans models for spheroidal systems, with distribution functions that depend on the exact integrals of motion: the energy, $E$, and the $z$ component of angular momentum, $L_z$. The distribution function for the disks depends on $E$, $L_z$, and a third "integral," $E_z$, the vertical energy, which is approximately conserved in cool stellar disks (see, e.g., Binney & Tremaine 1987). The radial velocity dispersion profiles correspond initially to a disk with Toomre (1963) $Q = 2.0$ at $R = 2.5R_d$. Evolved in isolation, the models experience no major transitions at startup, and they are stable against bar formation over the timescales of interest.

In our calculations, each galaxy is represented by 48,000 particles: 16,000 in the disk, 8,000 in the bulge, and 24,000 in the dark halo. Because we use a fixed number of halo particles, but vary the halo masses, individual halo particles have different masses from one run to another. As a result, the amount of disk heating due to two-body relaxation is more pronounced in the models with more massive halos. This effect is compounded by the fact that the more massive galaxies have a longer pre-collision evolution, as they are started further apart due to their more extended halos. Models of isolated galaxies allow us to quantify this disk heating: at the time of collision (when the tails are formed), Models A and B have $Q \approx 2.5$, while Models C and D have $Q \approx 4.0$ and 5.0, respectively. The relatively warm disks at the time of encounter caused concern about the validity of our results, particularly for the models with the most massive halos. To address this problem, we repeated two of the experiments with 5 times as many halo particles to reduce the growth rate of $Q$ and examine the sensitivity of our results to the disk heating. While the tidal features that develop in the large $N$, low $Q$ models are crisper than those in their high $Q$ counterparts, the morphologies and lengths of the tails are quite similar in both cases (see §§3 and 4.3 below), implying that our conclusions are insensitive to this aspect of the dynamics.

## 2.2. Orbital Parameters

Initially, we focus on exactly prograde mergers of equal mass disk galaxies from zero energy orbits. This choice of encounter parameters is motivated by our desire to match systems such as the Antennae or NGC 7252, which possess long tidal tails. As shown by Toomre & Toomre (1972), these features are most easily generated when comparable mass disk galaxies collide on a prograde orbit. The orbital energy is also chosen with tail-making in mind: galaxies on high speed unbound orbits will pass by one another too quickly to form long tails, while moderately bound orbits have encounter speeds only marginally slower than zero-energy orbits. Accordingly, we choose a zero energy orbit for our fiducial set of calculations, and consider bound orbits separately in §4.1. The galaxies are placed on their orbits with an initial separation $R$ chosen such that the dark halos are just touching; their relative velocity is then given by $\sqrt{GM_{tot}/R}$.

Given the orbital energy and disk inclinations, the remaining parameter to be fixed is the impact parameter, or pericentric distance of the initial orbit. The most appropriate choice here is less clear. While collisions with small impact parameters yield a stronger tidal perturbation, they are also faster, which may inhibit the tail-building process. Conversely, slower, more distant encounters have more time to raise tidal tails, but their tidal impulse will be weaker. Rather than fixing the impact parameter, we explore a range of possibilities with $b$=0.6, 1.2, 2.4, and 4.8 $R_d$. For a Keplerian orbit, $b$ is simply the perigalactic separation, $R_p$; however, galaxies are not point masses, and their extended mass distributions cause the orbits to diverge from a Keplerian trajectory. The exact value of $R_p$ and the relative velocity at periapse, $V_p$, will depend on the impact parameter $b$ and the mass distribution for the chosen model. (We found that the orbits of the galaxies until perigalacticon are well-traced by an orbit in the effective potential $W(r) = (M_2 \Phi_1(r) + M_1 \Phi_2(r))/2$ where $M_i$ and $\Phi_i(r)$ refer to mass and potential of each galaxy. While ad hoc, this potential predicts $R_p$ and $V_p$ within 10% for the simulations in this study.) The range of $(R_p, V_p)$ covered by our calculations is shown in Figure 2, where it can be seen that $R_p$ and $b$ are most discrepant for galaxies with the most massive, extended halos. In the discussion that follows we will refer to the models by their impact parameter $b$ rather than the varying perigalactic separation.

At first glance, it might appear as though the differences between $b$ and $R_p$ may make interpretation of the models difficult, since for a fixed $b$ we sample different values of $R_p$ in each of the models. In fact, since $V_p$ varies as well, this works to our advantage. For each value of $b$ the orbital *angular velocity* at periapse ($\Omega_{orb} = V_p/R_p$) is roughly constant in each of the models, as shown by the diagonal lines in Figure 2. The resonance between this orbital angular velocity and the rotational angular velocities ($\Omega_{rot}(R)$) in the disk is an important factor driving the formation of tidal tails during an interaction. Since the inner rotation curves of the galaxies are fixed, the different model disks all have identical $\Omega_{rot}(R)$, and collisions at fixed $b$ all sample similar values of $\Omega_{orb}/\Omega_{rot}(R)$, regardless of varying $R_p$ for the different models.

With four galaxy models and four impact parameters, our fiducial calculations involve a total of 16 different merger simulations. While idealized, these encounters provide a *best-case* situation for generating long tidal tails; if galaxies with massive halos do not develop extended tails under these conditions, it will be very difficult for them to form long tails under *any* conditions.

### 2.3. Numerical Techniques

All calculations were performed using a treecode (Barnes & Hut 1986; Hernquist 1987, 1990; Dubinski 1988), with a leapfrog timestep $\Delta t = 0.1$ (corresponding to $1.8 \times 10^6$ yr when scaled to physical units for the Galaxy) and a Plummer softening radius $\epsilon = 0.025$ (80 pc). We ran the simulations concurrently on the Pittsburgh Supercomputing Center CRAY T3D on a 16 node partition. The San Diego Supercomputing Center Paragon was also used for some preliminary calculations in a similar manner. Some simulations were also performed with a completely parallelized treecode (Dubinski 1995).

## 3. Fiducial Encounters

We begin by describing in detail the calculations involving collisions of galaxies with the lowest mass halos (Model A), then compare the evolution of the more massive models in turn. Figure 3 shows the evolution of the lowest mass halo galaxy mergers over a range of the parameter $b$. Shortly after the galaxies first collide (at $t \sim 45$), massive tidal tails form and begin to expand away from the merging galaxy pair. At the same time, secondary tidal features form (seen, e.g., extended from the galaxies at 90° from the primary tails at $t \sim 62$ in the $b = 0.6$ collision) from material in each disk which has passed through the companion galaxy. This material is much more diffuse than the primary tails, and may prove difficult to detect, but a possible example of such faint "anti-tails" can be seen in deep exposures of NGC 4038/39 (e.g., Schweizer 1978).

The primary tails form from material in each galaxy which is on the side of the disk away from the companion at perigalacticon, and whose velocity vector points largely away from the center of mass of the galaxy pair (see Figure 15 of Toomre & Toomre (1972) or Figure 8 of Hibbard & Mihos (1995)). The strong tidal forces from the companion galaxy, coupled with the motion of the particles away from the galaxies' center of mass, act to draw this material out into the massive, extended tidal tails. The respective companion acts to *add* kinetic energy to material in each tidal tail, and there is a one-to-one correspondence between the binding energy of material in the tails and its distance along the tail (Hibbard & Mihos 1995). Material at the base of the tail is most tightly bound to the galaxy pair, while the tips of the tails are composed of the outermost, least bound (and, in fact, sometimes unbound) tidal material. This distribution of energy along the tail results in rapid evolution in the structure of the tail — as the tips continue to expand, material in the base of the tail falls back into the galaxy pair, and the surface density of the tails rapidly declines (Mihos 1995).

The kinetic energy which goes into the tail material comes at the expense of orbital energy in the encounter. An even greater amount of orbital energy goes into the dark matter halos, as they are "spun up" by the encounter (Hernquist 1992, 1993b). The transfer of so much orbital energy to the internal motions of the galaxies results in a rapid braking of the galaxies on their orbits, and the pair quickly falls upon each other and merges into a single object displaying extended tidal tails. Violent relaxation in the inner regions of the remnant rapidly restructures the galaxies stellar components into an elliptical-like $R^{1/4}$ profile over a large range of radius (e.g., Negroponte & White 1983; Barnes 1988, 1992; Hernquist 1993c; Hernquist, Spergel, & Heyl 1993). As the remnant passively evolves from this point onwards, material from the tails continues to rain in on the remnant, at an ever-decreasing rate (Hibbard & Mihos 1995).

Several effects can be seen as the impact parameter $b$ (or, equivalently, the perigalactic separation $R_p$) is increased for the collisions. First and most noticeably, the galaxies take longer to merge. While the closest collisions result in a merger in $\sim 20$ time units ($\sim 400$ Myr), the more distant encounters are only merging after twice that time. As discussed by Farouki & Shapiro (1982) and Barnes (1992), orbital decay is strongly dependent on the ratio $R_p/R_{1/2}$, (where $R_{1/2}$ is the half mass ratio of the progenitor galaxies); for the encounters shown in Figure 3 $R_p/R_{1/2}$ ranges from 0.5 to 1.5. As dynamical friction is more efficient when the galaxies are deeply interpenetrating, at fixed orbital energy merging is more rapid for closer encounters.

A second, more subtle trend with $b$ can be seen in the morphology of the tidal tails. The tails formed from close collisions appear rather linear, while wider encounters produce tidal tails which are significantly more curved. This effect results simply from the varying amounts of angular momentum contained in the orbits of the galaxy pairs. Galaxies involved in the very close $b = 0.6$ collisions travel on low angular momentum, nearly radial orbits; material in the tidal debris *necessarily* travels on similar radial orbits, resulting in the linear tidal tails. In contrast, the wider $b = 4.8$ orbits are characterized by a much higher angular momentum, and the tails which form in these encounters also are populated by material on large angular momentum, curved orbits. In essence, therefore, the morphology of the tidal tails tracks the shape of the initial orbit of the merging galaxies. Unfortunately, this diagnostic will be difficult to use in practice to constrain the orbits of observed merging galaxies, as even the most curved tidal tail may appear quite linear when seen in projection.

Turning now to models with more massive and extended halos, Figure 4 shows the mergers of galaxies with halo:disk+bulge mass ratios of $\sim 8$ (Model B). Differences between these mergers and the lower mass, compact halo galaxy mergers shown in Figure 3 can easily be seen. At fixed impact parameter, these galaxies take longer to merge than their low mass counterparts, an effect due in large part to the faster encounter speed for these collisions — at perigalacticon, the relative velocity is $\sim 20\%$ faster for the higher mass galaxies.[5] At higher speeds, the transfer of orbital energy to internal motions of the galaxies

---

[5]Because the orbits diverge from pure Keplerian orbits when the galaxies interpenetrate, the simple

is less efficient (as $\Delta E \sim v^{-2}$; e.g., Binney & Tremaine 1987) and the orbit takes longer to decay.

More interestingly for our study is the differences in the structure of the tidal tails formed in these encounters. As the morphology and surface density of the tails evolves rapidly with time, we need to compare the various models *at similar evolutionary times.* Specifically, we examine the appearance of the tails at the point where the galaxies have just merged, in order to compare the tails to those observed in relatively young mergers like the Antennae, the Superantennae, and NGC 7252. The tails in these more massive model B galaxy mergers appear both lower in mass and less extended when compared to tails produced in the lower mass model A collisions.

This trend was anticipated by White (1982) and Negroponte & White (1983), and demonstrated explicitly by Barnes (1988). As the mass of the dark halos increases, the encounter velocity increases, shortening the encounter timescale and reducing the amount of energy imparted to the disks. The higher velocities also detune the resonance between the angular velocity of stars in the disks and the orbital angular velocity of the perturbing companion. While this resonance is very broad, as the angular velocities become more discrepant, the amount of energy imparted to material in the disk is reduced and tail-building is inhibited. At the same time, the potential well from which the tails must climb out of is deeper for the more massive halo, and so what tails are being produced have difficulty reaching large distances from the merging pair.

While the trend of weaker tails with increasing halo mass seems clear, the tails in these mergers are still quite prominent, and do not seem to conflict with the observed properties of tidal tails in young mergers. By way of example, observed at $t = 87$, shortly after merging, the tails produced in the Model B, $b = 2.4$ collision contain $\sim 6\%$ of the the luminous (disk+bulge) mass in the system and extend out to $\sim 35R_d$, or roughly $\sim 140$ kpc. These tails are comparable to the those of NGC 7252, which contain $\sim 7\%$ of the R-band light of the system and extend to $\gtrsim 80$ kpc (Hibbard 1994). However, the model B galaxy halos are still relatively low in mass when compared to some recent estimates of the halo masses of the Milky Way and other nearby spiral galaxies. If the trend of weakening tails continues to even more massive dark halos, then tails produced by mergers of galaxies like the Milky Way would prove very anemic indeed.

To examine this possibility, we next examine the tails formed by mergers of galaxies with even larger halo masses, a range of parameter space unexplored in previous studies. Figure 5 shows mergers of Model C galaxies, with a halo:disk+bulge mass ratio of 16, while

---

scaling of $v_{peri} \propto \sqrt{M_{tot}}$ does not apply.

Figure 6 shows the Model D mergers, with a halo:disk+bulge mass ratio of 30. These mass ratios are more in accord with the dark halo masses inferred for the Milky Way. As can be easily seen, the trends observed in the lower halo mass mergers continue quite dramatically at these halo masses. The galaxies take much longer to merge than their low-mass counterparts, and the tails continue to weaken. Even shortly after perigalacticon, when the tidal material is initially lifted, the tails are short and stubby, and they quickly fall back into their host galaxies long before the merger is complete. In fact, in the Model D mergers the tails resemble features more like open spiral arms than the classic extended tidal tails observed in many galaxy mergers. As these encounters were chosen to optimize tail-making (i.e. prograde, equal-mass, parabolic collisions) the fact that the tails are quite weak suggests that galaxies with massive halos would have difficulty forming tidal tails under *any* circumstances.

We were concerned that the structure of the tails might be affected by two-body relaxation in the more massive halo models. To test this, we repeated the $b = 2.4$, Model D merger with five times as many halo particles, reducing the artificial heating cause by the massive halo particles. The results of this test, included in Figure 6, show that while the tails are crisper and more well-defined when the numerical heating is reduced, they are no more massive or extended than in the fiducial models. Accordingly our results are relatively unaffected by discreteness noise in the models; equivalently, the results will scarcely be different for galaxies of (moderately) different values of the Toomre $Q$ parameter.

To examine the structure of the tails as the galaxies ultimately merge, we carried these simulations forward and illustrate a subset of the secondary encounters in Figure 7. The extended tidal features which are produced in the final merging are hotter and more dispersed than the thin tails of the initial encounter. They are more akin to the shells seen in nearly head-on mergers of disks (e.g., Hernquist & Spergel 1992). Part of the effect is that the secondary encounters do seem to be nearly head-on, as the dark halos absorbed most of the orbital angular momentum on the first encounter. The galaxies themselves are much hotter as well because of the strong agitation in the first encounter and this might partly account for the more diffuse tails.

In summary, the major trend is for mergers with more massive halos to have shorter tidal tails. Massive halos lead to faster encounters which weakens the perturbation on each disk. Figure 2 shows that the encounter velocities for the highest mass Model D are about twice that of the lowest mass model A for a given perigalactic separation. The velocity perturbations on the disk should be considerably weaker in the higher mass galaxy. Also, the deeper potential wells of high mass models reduce the maximum turnaround radius for a given perturbation velocity. Figure 8 illustrates this point by plotting the turnaround radius of orbits ejected from the disk at different radii as a function of an assumed purely radial velocity perturbation added to the circular velocity of the disk. We only show results for the extremes, Model A and D; the other models lie in between as expected. The potential of the galaxies are assumed to be unperturbed and isolated for this calculation so the numbers are fairly conservative and probably overestimate the true values. For Model A, a radial velocity perturbation equivalent to the disk rotational velocity, $V_r/V_c = 1.0$, is sufficient to eject material with $R > 3R_d$ to at least 30 scale lengths (120 kpc). The long tails in Figure 8 suggest that the velocity perturbation was therefore of order $V_r/V_c = 1.0$. An equivalent perturbation in Model D ejects material only to 10 scale lengths from the disk center before it falls back in. This predicted length agrees with the short tails seen in Figure 8 for the Model D merger which indeed only extend to 10 scale lengths. The perturbation again must be of order $V_r/V_c = 1.0$ but is probably smaller because of the shorter duration of the encounter. In Model D, a perturbation of $V_r/V_c = 2.0$ is required to eject material at the edge of the disk to 30 scale lengths and it appears that the high speed encounters of these high mass models cannot impart such a large perturbation.

In light of these results, then, we are left in somewhat of a quandary. While many of the most dramatic nearby examples of galaxy mergers show extended tidal tails, our calculations indicate that galaxies with massive halos produce only short, anemic tidal tails. If the Milky Way's halo is in fact as massive as suggested by the timing argument, then the halos of the observed merging galaxies must be much more compact and lower in mass than that of the Milky Way. We are left with a number of alternative conclusions, none of which seem ultimately satisfying. Either the structure of the Milky Way and other nearby spirals is unlike that of the majority of galaxies involved in interactions, or else the halo masses derived from the timing argument and/or satellite galaxy kinematics are overestimates. Before discussing the ramifications of these alternatives in any detail, we follow up on these fiducial models with a set of calculations which further explore parameter space and test the robustness of our initial results.

## 4. Other Encounters

Admittedly, the range of parameter space explored by our fiducial models is not comprehensive. While those models were chosen to favor the production of long tidal tails, it may be possible that unanticipated dynamics could make other types of encounters good candidates for tail-making. Two factors contribute to the weaker tails in the massive halo mergers: faster encounter speeds and deeper potential wells. While the potential well is fixed by the choice of halo model, several ways of slowing down the encounter and "re-tuning" the resonance between orbital and rotational angular velocity are possible. We now turn to some of these other encounters, focusing specifically on the massive halo galaxy models to see if these galaxies can form extended tidal tails under other conditions.

### 4.1. Bound Orbits

One possibility is that galaxies merge on bound, eccentric orbits rather than the parabolic orbits assumed in the fiducial models. A bound orbit would slow the encounter velocity at perigalacticon and may result in the formation of more prominent tails. While it is unlikely that any galaxy pairs which might form on tightly bound orbits would survive to the present time, it seems at least plausible that much wider encounters than those studied here may decay into a bound orbit for the second, much closer passage. Dynamical friction after this second passage would then result in a rapid merger, like those modeled in §3.

To explicitly test the evolution of tails in a bound encounter, we have set up a merger of Model D galaxies on a bound orbit. We choose a prograde orbit with $R_p = 3$ and eccentricity $e = 0.8$. At perigalacticon, the relative velocity of the galaxies is $V_p = 3.5$, considerably slower than the $V_p = 4.5$ expected for zero energy parabolic orbits. For comparison, the relative velocities for zero energy orbits in Models A and B at $R_p = 3.0$ are $V_p = 2.5$ and $V_p = 3.0$ (Figure 2). The results of this simulation are shown in Figure 9. The tidal tails which form initially, quickly fall back into the merging pair; whatever gain was achieved by slowing the interaction down was offset by the deep potential wells, and again the tails are quite anemic and very short-lived. Even slower, more tightly bound orbits, therefore, are not likely to improve the situation and, furthermore, seem astrophysically unreasonable.

### 4.2. Unequal Mass Ratios

A second way to reduce the encounter velocity is reduce the total mass of the galaxy pair. Simply scaling down the mass of *each* galaxy in lockstep will not help; the circular velocities of the disks will then also be reduced, leaving the ratio of orbital to rotational angular velocity unchanged. Instead, we change the *mass ratio* of the pair, slowing down the encounter while keeping fixed the circular velocity of one of the galaxies (hereafter referred to as the "primary"). However, while the slower encounter may help tail-making, the tidal field from the low mass companion is weaker than in an equal mass merger, and may completely offset the gain from the reduced encounter velocity.

To see which of these two effects dominates, we performed two calculations involving mergers of unequal mass galaxies, using 2:1 and 3:1 mass ratios. In each case, the low mass companion is constructed using galaxy Model D, scaled down in mass, and with a scale length derived from a $R_d \sim M^{1/2}$ relationship expected if the galaxies follow the $M \sim V_{circ}^4$ scaling implied by the Tully-Fisher relation. The primary galaxy is each merger is an unaltered Model D galaxy. The galaxies are then placed on a parabolic, prograde orbit with impact parameter $b = 2.5$ and an initial separation of $R_{sep} = 100$. The resulting perigalactic separation is $R_p \sim 5$ for each case.

Figure 10 shows the evolution of these two unequal mass mergers. The evolution proceeds in much the same way as the equal mass case, short spiral arms are thrown off each galaxy and wrap up before the second encounter. The second encounter is nearly head-on and hot tails and shells are thrown off instead of thin tails.

Several factors conspire to limit the tails produced in these unequal mass encounters. First, the reduced mass of the companion galaxy results in much weaker tidal forces acting on the primary in comparison to the equal mass mergers shown in §3. Second, although the encounter velocity is slowed somewhat in these encounters, it is not a great effect: $V_p = 3.5$ and 3.4, respectively, for the 2:1 and 3:1 mergers, compared to $V_p = 3.8$ for the comparable 1:1 merger. As the mass of the companion is reduced, the parabolic encounter speed simply drops asymptotically towards the escape velocity for the primary galaxy, $V_{esc} = 3.3$ at the given perigalactic distance, and not much time is gained for the encounter to strongly perturb the disks.

### 4.3. Inclined Disks: The Antennae

Yet another factor which can influence the development of tidal tails is the orientation of the disk plane to the orbital plane. Toomre & Toomre (1972) showed that tails were best formed during prograde collisions, but that highly inclined encounters were still effective at creating tails. However, Toomre & Toomre considered point-mass galaxies, which allowed for a good match between $\Omega_{orb}$ and $\Omega_{rot}$. In galaxies with extended mass distributions, like those considered here, the higher relative velocities at impact appear to inhibit tail formation in prograde encounters. In principle, the situation could be somewhat better for inclined encounters, as the orbital angular velocity of the perturber *projected onto the disk plane* is reduced in such encounters. One should note, however, that although the encounter may be more resonant, the duration of the encounter is about the same for different disk orientations and this may be the controlling factor.

To test this case using a specific example, we attempted to reproduce the well-studied and often-modeled galaxy merger "The Antennae" (NGC 4038/39) using our four different galaxy models. Barnes (1988) previously modeled this system with fully self consistent galaxies with halo:disk+bulge mass ratios of 4:1, and was able to reproduce much of the tidal tail morphology of the galaxies. Our new simulations extend his efforts to include galaxies having much more massive halos. For each of the different galaxy models, we set up a merger using Toomre & Toomre's (1972) disk orientations for the encounter. The galaxies are each inclined to the orbital plane by $i = 60°$, with arguments of pericenter $\omega = -30°$. The galaxies are again placed on zero energy orbits, with impact parameters $b = 2.5$. On these orbits, $R_p \sim 4$ for the collisions, somewhat smaller than Barnes' (1988) choice of $R_p = 6$. However, the differences between our Model A merger and Barnes' model for the Antennae proved relatively minor.

Figure 11 shows the four models projected onto the orbital plane around the times when the galaxies exhibit the longest tidal tails. Models A and B closely resemble Barnes' (1988) two simulations with 4:1 and 8:1 halo:disk+bulge mass ratios. Models C and D again show the difficulty in producing tidal tails in the high speed encounters resulting from the massive halos of these galaxy models. As before, the tails extend only to $\sim 10$ scale lengths before quickly falling back into the galaxies well before the they actually merge. Tails produced in the subsequent merging are even shorter and less massive than those shown here (c.f. Figure 6).

The failure of Models C and D to reproduce the Antennae is further emphasized by viewing the simulation from other directions. For each simulation, lines of sight were chosen such that in projection the galaxies looked most like the observed morphology of the Antennae. Figure 12 shows the view in the orbital plane down lines of sight 80°, 70°, 60°, and 50° from the line of pericenter for Models A, B, C, and D, respectively. Models A and B exhibit the long, thin, curving tidal tails for which the Antennae is famous, but as the halo mass is increased, as in Models C and D, the tails are more like low mass, stubby "plumes" than the tidal tails of the Antennae. Concerned again about the consequences of greater disk heating in the massive halo models, we reran the Model D Antennae using five times as many halo particles to reduce this disk heating. Although the "tails" are somewhat thinner and more well-defined (Figure 12), they are still extend to short distances when compared to both the low mass halo models and the Antennae itself. This model demonstrates again that these numerical affects do not significantly alter our major conclusions.

### 4.4. A More General Case: The Milky Way and Andromeda

So far, our calculations represent a controlled search through parameter space to examine how halo masses affect the properties of tidal tails in galaxy pairs, and we find generically that galaxies with massive halos produce short and low mass tails upon merging. As a final test of our results, we look at a more general case of a galaxy merger with unequal masses, inclined geometries, and a bound orbit. Using the structural and kinematic properties of the Milky Way and the Andromeda Galaxy, we present a possible merging scenario to explore whether the future merger of the Milky Way/Andromeda system will result in the spectacular tails seen in the Antennae, or merely a dull, amorphous merger.

To set up the merger, we must first adopt mass models for the Milky Way and Andromeda. A variety of arguments suggest that the mass of Andromeda, $M_A$, is approximately twice that of the Milky Way, $M_{MW}$. The flat portion of Andromeda's rotation curve has $V_A = 260$ km s$^{-1}$ (e.g., Braun 1991), and the B luminosity disk scale length is $R_A = 5.8 \pm 0.3$ kpc (Walterbos & Kennicutt 1988). These values can be compared to those for the Milky Way, $V_{MW} = 220$ km s$^{-1}$ and $R_{MW} = 4.5 \pm 1.0$ kpc (Freeman 1987). Assuming that Andromeda is a scaled-up version of the Milky Way, $M_A/M_{MW} = (R_A V_A^2)/(R_{MW} V_{MW}^2) \approx 1.9$. The Tully-Fisher relation, $M \propto V^\alpha$ with $\alpha = 3 - 4$, also predicts $M_A/M_{MW} \approx 1.5 - 2$, (e.g., Raychaudury & Lynden-Bell 1989).

Having fixed the mass ratio of the galaxies at $M_A/M_{MW} = 2$, we now need to set the *total mass* of the galaxy pair. Various arguments suggest that the total mass of the Milky Way and Andromeda is quite large. Perhaps most convincingly, given the separation $D = 700$ kpc and radial velocity $v_r = -130$ km s$^{-1}$, the timing argument gives a total mass for the system of $M_A + M_{MW} = 4.5 \times 10^{12}$ M$_\odot$ for a Universe age of 13 Gyr (i.e. $H_0 = 50$ km/s/Mpc, $\Omega = 1.0$ or $H_0 = 80$ km/s/Mpc, $\Omega = 0.2$). This sets the mass of the Milky Way at $M_{MW} = 1.5 \times 10^{12}$ M$_\odot$, similar to our Model D galaxy. For the Milky Way, we therefore adopt the scaling parameters $R_d = 4.0$ kpc and $V_c = 220$ km s$^{-1}$ for model D giving a total mass, $M_{MW} = 1.7 \times 10^{12}$ M$_\odot$. For Andromeda, we adopt, $R_d = 6.0$ kpc and $V_c = 260$ km s$^{-1}$ giving a mass $M_A = 3.6 \times 10^{12}$ M$_\odot$. We also set up a low mass foil of this case using Model B as the base model. With the same distance and velocity scalings the masses in this case are $M_{MW} = 5.0 \times 10^{11}$ M$_\odot$ and $M_A = 1.1 \times 10^{12}$ M$_\odot$, values in accord with the lower bound of the mass estimates of the Milky Way.

The final constraints on the merger are the orientations, separation, and relative velocity of the galaxies. In Galactic coordinates, the spin axis of Andromeda points in the direction $(l, b) = (270°, -30°)$ (Raychaudury & Lynden-Bell 1989) while the Milky Way spin axis points to the south galactic pole, $(l, b) = (0°, -90°)$ by definition. We adopt a separation $D = 700$ kpc with Andromeda currently positioned at $(l, b) = (121°, -23°)$ and a radial velocity $v_r = -130$ km s$^{-1}$. Andromeda's transverse velocity $v_t$ is unknown; however, in keeping with the spirit of this work, we choose $v_t$ such that the galaxies can best build tidal tails. For Model D galaxies, we use $v_t = 26$ km s$^{-1}$, pointing towards $(l, b) = (180°, 0°)$ so the resulting orbit has $R_p = 20$ kpc. For Model B, we use $v_t = 20$ km s$^{-1}$ which also leads to $R_p = 20$ kpc. In these encounters, the disk of the Milky Way is inclined 23° to the orbital plane, as closely aligned to prograde as the constrained properties allow. The first part of the orbit is uneventful so we advance the galaxies along a Keplerian orbit with the given initial conditions until they are separated by 400 kpc and begin the N-body simulation at this time. From this point, the time to impact is only $\sim 1$ Gyr for both cases.

Figure 13 presents the time sequence of the interactions for the low and high mass models from a view looking down the North Galactic Pole. The orbital plane is inclined slightly with respect to this line of sight. The smaller circular galaxy in the initial frame of each sequence is the Milky Way. Our expectations from the previous simulations of high mass models are born out yet again. In the low mass models, the encounter is very resonant and long tidal tails are thrown off from both galaxies. Material is effectively ejected from the Milky Way, and merging occurs shortly after the first pass. In the high mass models, the encounter velocity is too high on the first pass to lead to a strong resonant response in either galaxy. Short tidal tails (5 scale lengths) are thrown off both galaxies but quickly fall back in. The secondary encounter which leads to the final merger also fails to develop large tidal tails. Like the other experiments, the galaxies fall together on a nearly radial orbit on the second encounter and merge together without throwing off any significant tidal tails. It appears, therefore, that if the large median mass estimates of the Milky Way's halo from the timing argument and satellite kinematics are correct, then the halo:disk+bulge mass ratio of our Galaxy must be approximately 3 times that in galaxies with long tidal tails such as the Antennae or Superantennae. On the other hand, the merger of a low mass Galaxy with $M = 5 \times 10^{11}$ M$_\odot$ with a similar low mass Andromeda *will* resemble the Antennae. A low mass Model B Galaxy is still consistent with the lower bounds on the mass estimates from satellite kinematics.

## 5. Discussion

Our calculations demonstrate that mergers of galaxies with massive, extended dark matter halos result in short-lived, very anemic tidal tails. In contrast, many observed galaxy mergers display tidal tails which extend for $\sim 50 - 100$ kpc, and contain as much as 20% of the system luminosity (Hibbard 1994). Taken together, these results suggest that these observed merging galaxies must have relatively compact, low mass halos, with halo:disk+bulge mass ratios of $\lesssim 10$:1. This conclusion is at odds with several different observational and theoretical studies of dark matter halos, which suggest that the halos around spiral galaxies extend to several hundred kpc and have masses $\gtrsim$ a few $\times 10^{12}$ M$_\odot$, implying halo:disk+bulge mass ratios of $\gtrsim$ 20:1. How, then, can these different results be reconciled with one another?

One possibility is that the current determinations of halo masses of spirals may be overestimates. The strictest constraints on the mass distributions in spiral galaxies come from optical and HI rotation curves. Unfortunately, these techniques can probe the mass distribution out to $\lesssim 30 - 40$ kpc. At larger radii, our only constraints come from the Local Group timing argument and the kinematics of satellite galaxies. Each of these techniques is subject to considerable uncertainty. Mass estimations based on satellite galaxy kinematics are sensitive to the shapes of the assumed satellite orbits, and furthermore, are confined to relatively small sample sizes. Zaritsky & White (1994) have attempted to circumvent these issues by combining satellite populations around galaxies of similar luminosity and Hubble type, and using numerical infall models to constrain the orbits of the satellites. Their results suggest a 90% confidence interval for the mass within 200 kpc of 1.5–2.6 $\times 10^{12}$ M$_\odot$, with more recent data suggesting halos which are even more massive and extended (Zaritsky, private communication).

The Local Group timing argument represents a simpler, and therefore perhaps more powerful, constraint on the mass of the Milky Way. Given the observed separation and radial velocity of the Milky Way/Andromeda pair, the total mass of the system is $\sim 5 \times 10^{12}$ M$_\odot$ for a 15 Gyr old Universe. This result is sensitive to both to the age of the Universe and to the orbital evolution and accretion history of the Milky Way/Andromeda pair (e.g., Peebles et al. 1989). Kroeker & Carlberg (1991) used cosmological simulations to test the accuracy of timing argument masses, and found that galaxy masses were typically *underestimated* by a factor of two when using this approach, worsening the discrepancy between our results and those from the timing argument. An older Universe would reduce the derived mass from the timing argument, but given recent estimates of a relatively large Hubble constant (Freedman et al. 1994), this would require a non-zero cosmological constant $\Lambda$. In any case, increasing the age of the Universe to 20 Gyr only reduces the derived timing argument mass to $\sim 3.5 \times 10^{12}$ M$_\odot$.

A second possible conclusion is that we are seeing evidence for real, physical differences between the halos of nearby spirals and those of prominent mergers like the Antennae. There may simply be a distribution of halo masses and extents for a given galaxy luminosity, and the fact that the most famous examples of mergers seem to have low mass halos is merely a selection effect resulting from the fact that such galaxies produce the most dramatic tidal tails. Alternatively, local environment may play a role in determining the halo properties of a galaxy. Galaxies forming in low density regions of the Universe may have smaller halos if accretion stops at early times – essentially such galaxies live in a locally low $\Omega$ universe — while galaxies in higher density regions such as groups and clusters may have more massive, extended halos. It is true that the Antennae, Superantennae, and NGC 7252 are all field (rather than cluster) objects; unfortunately, countering this argument is the satellite galaxy analysis of Zaritsky & White (1994) – although the galaxies in their sample were selected to be isolated, the "composite" derived halo mass was still $\gtrsim 10^{12}$ $M_{\odot}$. In addition, dynamical evolution in clusters may further cloud any signature of the initial conditions, as extended halos may be partially stripped away over a Hubble time.

Our results are also in disagreement with theoretical expectations of galaxy halo formation in CDM cosmologies. Continued infall in $\Omega = 1$ cosmologies should result in very massive galaxy halos (Gunn & Gott 1972), an expectation verified through various numerical simulations (e.g., Zurek et al. 1988, Navarro et al. 1995). In fact, the recent simulations of Navarro et al. suggest that CDM dark matter halos should extend beyond 200 kpc, with total masses well in excess of $10^{12}$ $M_{\odot}$. In low $\Omega$ universes, however, accretion stops at an earlier redshift, and galaxy halos may be more in accord with the lower mass, more compact halos suggested by our tidal tail models (e.g., Xu 1995). Alternatively, mixed (cold + hot) dark matter cosmologies predict lower galaxy halo masses than do pure CDM models (e.g., Klypin et al. 1993), and may also be consistent with the constraints set by the tidal tails.

It appears that there is no simple resolution to the problems discussed above. However, the argument put forward here, that one can use tidal features in merging systems to constrain the mass distribution in galaxies, is convincing because of its simplicity. By considering a wide range of encounters, we have eliminated uncertainties arising from orbital parameters. Assuming that we understand the distribution of luminous matter in disks, the only remaining parameters are those defining the structure of the dark halos. These final unknowns are further constrained at small radii by observations of galaxy rotation curves. Here, we have chosen to focus on models whose rotation curves within the Holmberg radius are reasonable caricatures of those in real spirals. The remaining uncertainty, therefore, resides in the asymptotic properties of the halos, as embodied by their extents and total masses. Unless we have overlooked something fundamental, the simulations presented here provide a very simple and robust prescription for studying the global properties of halos in galaxies.

The implications of our results for theories of structure formation on even larger scales are less clear–cut. Since the data set we are comparing our findings to is admittedly small, it is always possible to argue that the observed systems are for some reason not representative of most galaxies. To the extent we can extrapolate our results to galaxies generally, the simulations reported here provide tantalizing evidence that halos are significantly more compact and less massive than those expected in CDM cosmologies with $\Omega = 1$, arguing perhaps for a lower density universe.

We thank Scott Tremaine, Josh Barnes, Alar Toomre, Sandy Faber, and Dennis Zaritsky for several helpful discussions and constructive suggestions during the course of this study. This work was supported in part by the Pittsburgh Supercomputing Center, the San Diego Supercomputing Center, the Alfred P. Sloan Foundation, NASA Theory Grant NAGW–2422, the NSF under Grants AST 90–18526 and ASC 93–18185 and the Presidential Faculty Fellows Program.

Table 1: Galaxy Model Properties

| | Disk | | | Bulge | | Halo | |
|---|---|---|---|---|---|---|---|
| Model | $M_d$ | $\sigma_{r,0}$ | $R_t/R_d$ | $M_b$ | $R_b/R_d$ | $M_h$ | $R_h/R_s$ |
| | (1) | (2) | (3) | (4) | (5) | (6) | (7) |
| A | 0.82 | 0.47 | 6.0 | 0.42 | 1.0 | 5.2 | 21.8 |
| B | 0.82 | 0.47 | 6.0 | 0.42 | 1.0 | 9.6 | 30.1 |
| C | 0.82 | 0.47 | 6.0 | 0.42 | 1.0 | 19.8 | 44.0 |
| D | 0.82 | 0.47 | 6.0 | 0.42 | 1.0 | 37.0 | 72.8 |

Note. — (1) disk mass, (2) disk central radial velocity dispersion, (3) disk radial extent (radius where density drops to zero), (4) bulge mass, (5) bulge radial extent, (6) halo mass, (7) halo radial extent.

– 23 –

– 26 –

Fig. 1.— Rotation curves for the model galaxies. a) Inner rotation curves, b) Outer rotation curves

Fig. 2.— Velocity, $V_p$ versus separation, $R_p$ at perigalacticon for zero energy orbits of prograde, equal mass galaxy mergers for the four models. The points refer to the trajectories of the 16 simulations described in the text. Lines of constant angular velocity, $\omega$, at various radii in the disk are also plotted. The intersection with curves of $V_p$ vs. $R_p$ show where the orbital angular frequency is resonant with various disk spin frequencies for different galaxy trajectories at perigalacticon.

Fig. 3.— Mergers of Model A galaxies. Simulations are parameterized by their impact parameter $b$. Time progresses downward, beginning at the time of the collision. Time is measured in simulation units where 1 unit=0.18 Gyr scaled to the Milky Way. The width of each snapshot is 80 $R_d$ or 320 kpc scaled to the Milky Way.

Fig. 4.— Mergers of Model B galaxies. Arrangement and scaling as in Figure 3.

Fig. 5.— Mergers of Model C galaxies. Arrangement and scaling as in Figure 3.

Fig. 6.— Mergers of Model D galaxies. "Large N" refers to a rerun of the $b = 2.4$ merger with 5 times as man halo particles (see text). Arrangement and scaling as in Figure 3.

Fig. 7.— Secondary encounters and mergers of selected Model C and D galaxies. Scaling as in Figure 3.

Fig. 8.— Turnaround radius, $R_a$ versus radial velocity perturbation, $V_r$ at various radii in the disk for Models A and D.

Fig. 9.— Merger of Model D galaxies on a bound orbit.

Fig. 10.— Mergers of unequal mass Model D galaxies.

Fig. 11.— Models of NGC 4038/9, the Antennae, as viewed in the orbital plane. Time is measured relative to the point of impact in simulation units. The width of each snapshot is 30 $R_d$ or 120 kpc scaled to the Milky Way.

Fig. 12.— Models of NGC 4038/9, the Antennae, viewed along directions to resemble the observed galaxy. Lines of sight are in the orbital plane and lie 80°, 70°, 60°, and 50° from the x-axis (or line of periapse for the chosen orbit) for each of the Models A, B, C and D respectively. The width of the snapshot is 24 $R_d$ or 96 kpc scaled to the Milky Way.

Fig. 13.— Merger of the Milky Way and Andromeda galaxies using low and high mass models for the system. All observed constraints are used to set up the orbit though the unknown transverse velocity is chosen to give a close, nearly prograde encounter to maximize the response in the disks. The low mass model produces long tidal tails while the high mass model fails in this regard, following the trend of the other simulations.